\documentclass[%
 reprint,
superscriptaddress,
showpacs,preprintnumbers,
 amsmath,amssymb,
 aps
]{revtex4-1}

\usepackage{graphicx}
\usepackage{dcolumn}
\usepackage{bm}
\usepackage{soul}
\usepackage{color}
\newcommand{\uvec}[1]{\boldsymbol{\hat{\textbf{#1}}}}

\begin{document}


\title{Trapping flocking particles with asymmetric obstacles}

\author{Raul Martinez}
\affiliation{Departamento de F\'isica Te\'orica de la Materia Condensada, Instituto Nicol\'as Cabrera
and Condensed Matter Physics Center (IFIMAC), Universidad Aut\'onoma de Madrid, E-28049 Madrid, Spain}
 \affiliation{Departamento de Estructura de la Materia, F\'isica T\'ermica y Electr\'onica, Facultad de Ciencias F\'isicas, Grupo Interdisciplinar de Sistemas Complejos, Universidad Complutense de Madrid, 28040 Madrid, Spain.}
\author{Francisco Alarcon}%
\affiliation{Departamento de Estructura de la Materia, F\'isica T\'ermica y Electr\'onica, Facultad de Ciencias F\'isicas, Grupo Interdisciplinar de Sistemas Complejos, Universidad Complutense de Madrid, 28040 Madrid, Spain.}
\affiliation{Departamento de Ingenier\'i­a F\'i­sica, Divisi\'on de Ciencias e Ingenier\'i­as, Universidad de Guanajuato, Loma del Bosque 103, 37150 Le\'on, Mexico }
\author{Juan Luis Aragones}
\email{Co-corresponding Author. E-mail: juan.aragones@uam.es}
\affiliation{Departamento de F\'isica Te\'orica de la Materia Condensada, Instituto Nicol\'as Cabrera
and Condensed Matter Physics Center (IFIMAC), Universidad Aut\'onoma de Madrid, E-28049 Madrid, Spain}
 \author{Chantal Valeriani}%
 \email{Corresponding Author. E-mail: cvaleriani@ucm.es}
  \affiliation{Departamento de Estructura de la Materia, F\'isica T\'ermica y Electr\'onica, Facultad de Ciencias F\'isicas, Grupo Interdisciplinar de Sistemas Complejos, Universidad Complutense de Madrid, 28040 Madrid, Spain.}
  
\date{\today}

\begin{abstract}
Asymmetric obstacles can be exploited to direct the motion and induce sorting of run-and-tumbling particles.
In this work, we show that flocking particles which follow the Vicsek model aligning rules experience a collective
trapping in the presence of a wall of funnels made of chevrons, concentrating at the opposite side of a wall of 
funnels than run-and-tumbling particles. Flocking particles can be completely trapped or exhibit a dynamical
trapping behaviour; these two regimes open the door to the design of a system with two perpendicular flows
of active particles. This systematic study broaden our understanding about the emergence of collective motion
of microorganisms in confined environments and direct the design of new microfluidics devices able to control
these collective behaviours.
   
\end{abstract}

\pacs{Valid PACS appear here}
\maketitle


\section{\label{sec:level1}Introduction}
Self-generated motion is ubiquitous in nature; microorganisms and cells are required to move to execute vital
biological processes. The dynamical behaviours emerging at the microscopic scale from this non-equilibrium nature
are fascinating~\cite{ramaswamy2010mechanics}. Recently, our attention has shifted towards not only  the effect of solid
interfaces on the dynamical behaviour of these active  systems~\cite{gomez2016dynamics}, but also  the
effect of their dynamics on their surroundings~\cite{patteson2018propagation, rabodzey2008mechanical}.

One of the most paradigmatic active matter systems are bacteria, whose locomotion mechanism~\cite{patteson2015running}
and collective behaviour~\cite{bozorgi2011effect} are highly influenced by the properties and structure of the supporting media. 
Some bacteria such as \textit{E. Coli} move following a run-and-tumble protocol, which alternates periods of swimming
in a straight line at constant speed with tumblings in which the movement orientation is randomly chosen, resulting 
into a diffusive dynamics, in the absence of external fields. However, the presence of asymmetric walls, such as
V-shaped obstacles, referred to as chevrons, produce  rectification of  bacterial motion, resulting into directed
motion~\cite{kaiser2012capture,berdakin2013influence}. It has  been recently shown that this ratchet effect can be utilised 
to extract work from a bacterial suspension  by means of the motion of anisotropic mobile obstacles such as  
gears~\cite{sokolov2010swimming,di2010bacterial,angelani2009self} or chevrons~\cite{a2010g,m2014c,s2015s,k2014t}.
Similarly, anisotropic fixed obstacles such as a wall of chevrons, with funnels between neighbouring obstacles,
 induce an increase of bacteria 
concentration on the concave side of the funnelled
wall~\cite{galajda2007wall,galajda2008funnel,hulme2008using,hulme2008using}.
This is a consequence of the persistence length of the bacteria movement, which increases the probability of crossing
the funnel from the convex to the concave side of the funnelled wall, effectively concentrating bacteria on the latter~\cite{wan2008rectification,reichhardt2017ratchet}.

Anisotropic obstacles are not only capable to induce ratchet effects on suspensions of run-and-tumbling particles, 
but also affect the collective behaviour of active aligning particles, 
such as Vicsek 
particles~\cite{vicsek1995novel}. 
This model, despite the simplicity of the inter-particle interaction rule, provides a 
successful platform to study self-organisation of active
particles~\cite{gregoire2004onset,chate2008collective,ginelli2016physics}, describing flock
formation and collective behaviour~\cite{toner2005hydrodynamics,vicsek2012collective}. 
In the presence of anisotropic obstacles such as a funnelled walls, Vicsek particles concentrate on one side of the wall~\cite{drocco2012bidirectional}. In this system, the  
rectification's direction can be reversed by changing  inter-particle steric-repulsion radius. 
In addition, it has been suggested that the alignment interactions between active particles interacting via alignment rules and  obstacles play a key role in rectifying active particles' motion in the presence of a wall of funnels
in an infinite periodic channel in which particles dynamics  obeys an over-damped Langevin rule with finite angular time correlation~\cite{zhu,hu}.

In the present work, we investigate the behaviour of active aligning particles interacting with arrays of chevron-like
obstacles that form parallel walls of funnels with alternating orientations, as shown in Fig.~\ref{comparison}A. 
Vicsek particles spontaneously accumulate on the convex  side of the funnelled walls. 
We study the dynamics of this collective self-trapping behaviour and identify two regimes, one static
and other dynamical in which particles are effectively trapped by constantly escaping and getting into the channels.
Exploiting these two regimes, we engineer a system  showing two perpendicular flows of Vicsek particles. 

\section{Methods}

To model active aligning particles we use the two-dimensional Vicsek 
model~\cite{gregoire2004onset,chate2008collective}, in which point-like particles align with their first neighbours' average 
orientation. Vicsek particles move at a constant speed of $v$ = 0.2, while their direction of motion, $\theta_i$, determines 
their velocity $\vec{v}_i=v(\cos (\theta_i) \uvec{x} + \sin (\theta_i) \uvec{y}$). The position of each Vicsek particle, 
$\vec{x_i}(t)$, is updated at regular time intervals $\Delta t$ according to:

\begin{align}
\theta _i (t+\Delta t) &  = \mathrm{arg}\left[\sum_{j\in \mathcal{N}_i(t)} e^{i\theta_j(t)}\right] + 2\pi\eta\zeta \label{upvel} \\
\vec{x_i}(t+\Delta t) & = \vec{x_i}(t)+\vec{v_i}(t+\Delta t)\Delta t
\label{uppos}
\end{align}
where $\eta\in [0,1]$ is the noise amplitude, $\zeta\in [-0.5,0.5]$ is a delta-correlated white noise and $\mathcal{N}_i$
is the list of first neighbours of particle $i$, which are defined as the particles inside a cut-off radius, R, from  particle $i$.
We refer to this scheme as metric Vicsek model~\cite{gregoire2004onset,chate2008collective}. Alternatively, we consider
Vicsek particles that only interact with their first Voronoi neighbours independently of the distance between them, referred to
as topological Vicsek model~\cite{ginelli2010relevance}. In our simulations, we set $R$ = 1 and $\Delta t$ = 1 and normalise
all quantities by $R$ and $\Delta t$.

We consider $N$ = 2000 Vicsek particles within a two-dimensional simulation box of length $L$ = 32 at constant 
density $\rho = N / L^2$ = 1.953. We apply periodic boundary conditions in both directions. For this system, the
order-disorder transition occurs at noises below $\eta$ = 0.5~\cite{gregoire2004onset,martinez2018}; thus, we only
consider noises ranging between 0 and 0.5, where flocking is observed. Besides Vicsek, we also study a system of 
run-and-tumble particles without aligning interactions. When considering run-and-tumble particles, we neglect the
first term in Eq.~\ref{upvel} and set $\eta$ = 1. The speed is set to 1 and so the persistence length. For a direct 
comparison with the Vicsek particles simulations, we consider a system with the same density  ($L=32$, $N=2000$). 

\begin{figure}
\centering
\includegraphics[width=\linewidth]{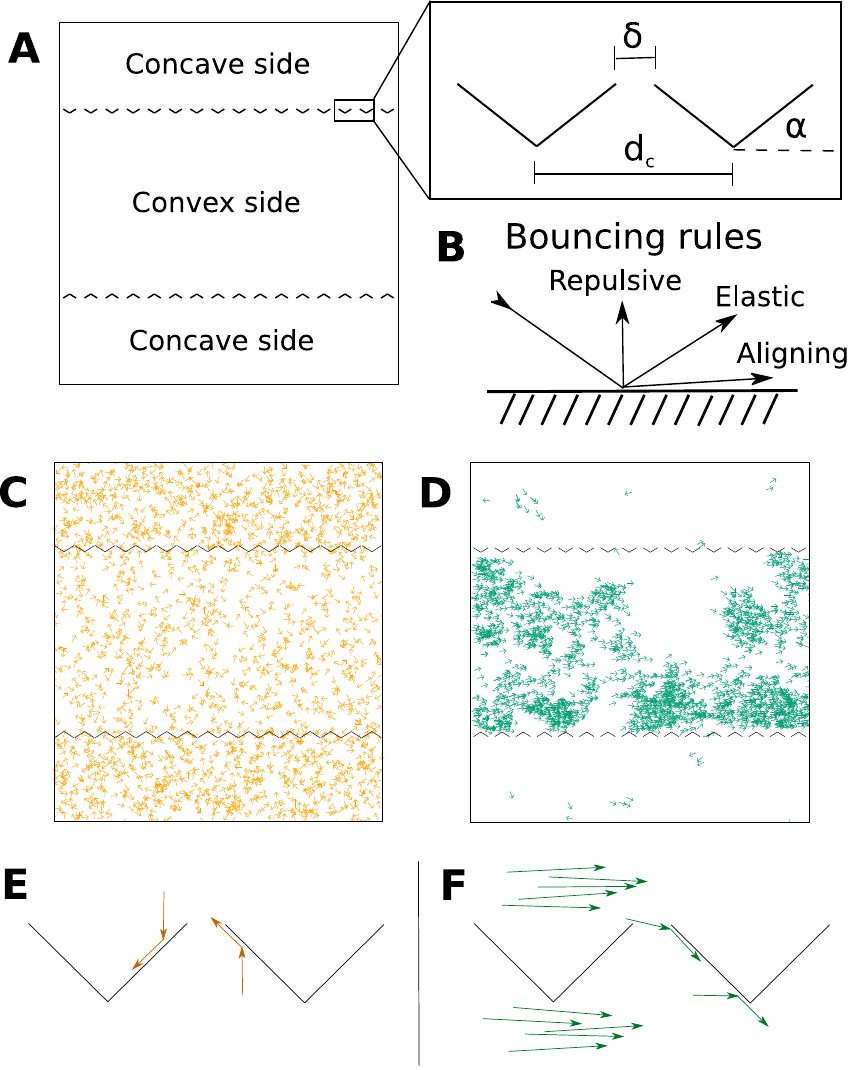}
\caption{
{\bf Scheme of the system.} A) Geometry of the channel made of walls of funnels formed by pairs of chevrons
of angle $\alpha$ with an opening $\delta$ and separated by a distance $d_c$. B) Scheme of the bouncing 
rules of Vicsek particles with the considered obstacles. Snapshots of the steady state reached by run-and-tumbling
particles (C) at $v=1$, $\delta$ = 0.1 and by Vicsek particles (D) at $v$=0.2 and $\eta$ = 0.1 swimming between two
parallel walls of funnels with $\delta$ = 1, $d_c$ = 2 and $\alpha$ = 30 degrees. Schemes showing the possible 
collisions between active particles and a funnel: E) for run-and-tumbling particles the probability  to pass
from the wide   towards the narrow opening side is high, while F) for a flock of Vicsek particles aligned with
the wall, collisions with the narrow opening side of the funnels force particles towards the wide opening side and
collision on the wide opening side only bounce them back.   
}
\label{comparison}
\end{figure}

In our system, active particles interact with fixed anisotropic obstacles with a chevron shape (V-shape). Details about
their implementation can be found in the SM. Briefly, when an active particle encounters a chevron it bounces, thereby 
reorienting its direction of motion. We consider three different bouncing rules: i) repulsive, ii) elastic and iii) aligning, 
schematically shown in Fig.~\ref{comparison}B. The repulsive bouncing rule is similar to that of 
Ref.~\cite{drocco2012bidirectional}, under which an active particle bouncing with a chevron changes its orientation 
to be perpendicular to the wall. According to the elastic bouncing rule, active particles are scattered from the 
chevrons according to the Snell's law; the approaching angle towards the wall and the outgoing angle from the
wall are the same. The aligning bouncing rules accounts for the tendency of the active particles to align with solid 
interfaces~\cite{galajda2007wall}. Active particles colliding with a solid wall change their orientations to align with it
independently of the approaching angle, plus a small angular random noise in the interval $[0, \eta \pi]$, being 
$\eta$ the noise for Vicsek particles or $\eta$ = 1/32 for run-and-tumble particles~\cite{galajda2007wall}.
To compute the time evolution of the particles that bounce with solid walls we use an
event-driven dynamics, whereas Vicsek update rules are implemented at regular intervals. Details about the
implementation of the models, interactions with obstacles and dynamics can be found in the supplementary material (SM). 

\section{Results and Discussion}

To start with, we study the behaviour of run-and-tumble and metric Vicsek particles in the presence of two parallel 
funnelled walls made of chevrons, where the walls are oppositely oriented, as shown in Fig.~\ref{comparison}A.
This geometry results in two regions, one on the concave and the other on the convex side of the funnels. 
Once reached the steady-state, both types of particles experience a trapping behaviour induced by the anisotropy 
of the obstacles' arrays. However, while run-and-tumble particles concentrate on the concave side of the funnels
(Fig.~\ref{comparison}C), Vicsek particles are trapped on the convex side (Fig.~\ref{comparison}D). 
These different trapping behaviours result from their differences in the interactions with the obstacle arrays. 
Run-and-tumble particles tend to go from the convex to the concave side of the funnelled wall due to their alignment 
with the chevrons's edges, as schematically shown in Fig.\ref{comparison}E. Thus, to induce trapping, the opening
between chevrons must be small compared to the run's persistence length~\cite{wan2008rectification}. On the contrary,
Vicsek particles are confined in the region between the convex sides of the funnelled walls. Differently from the trapping
experience by run-and-tumbling particles, the trapping of Vicsek particles is of collective nature. Initially, Vicsek particles
align with each other forming flocks. Next, flocks align their direction of motion with the funnelled walls, independently of
the chosen bouncing rules (see SM). When flocks are on the concave side of the walls of funnels, the particle's collisions
with the chevrons push them towards the convex side. Alternatively, the particles of flocks on the convex side of a funnelled
wall are bounced back, as represented in Fig.~\ref{comparison}F. Therefore, for a flock of particles aligned with a funnelled
wall, the probability of passing from the concave to the convex side is much higher than the other way around (see SM).
Since this trapping of Vicsek particles on the convex side of the funnelled walls is due to the interaction with the chevrons
of previously formed and wall-aligned flocks, the aligning interactions between Vicsek particles on the concave side and 
those on the convex side of the wall are not driving this trapping behaviour. We observe the same steady-state in systems
where Vicsek particles are not allowed to interact through funnelled walls (data not shown).

This collective trapping behaviour on the convex side of the funnelled walls clearly differs from the one described in
Ref.~\cite{drocco2012bidirectional}, where point-like Vicsek particles are preferentially located on the concave side of
the wall. Both systems differs in the chevrons' geometry, particles'  swimming speed and that we consider periodic
boundary conditions instead of a finite simulation box bounded by repulsive walls. A detailed comparison can be found in
the SM. However, the different trapping behaviours observed are consequence of the fact that both rectification mechanisms
are of different nature. In our system the rectification is of collective nature; it is the flock aligning with the funnelled walls 
what enables the rectification, which results into chevrons directing the particles towards the convex side of the wall,
as depicted in Fig.~\ref{comparison}F. Whereas in the system of Ref.~\cite{drocco2012bidirectional}, in which the flocks
do not align with the funnelled walls due to the chevron's angle and the presence of repulsive boundaries of the simulation
box, the rectification of Vicsek particles is similar to the one experienced by run-and-tumbling particles on the concave side
of the wall of funnels. Therefore, one can switch between rectification mechanism, and thus trapping localisation, 
by promoting or neglecting flock's alignment with the funnelled walls. For example, one can increase the angle of the
chevrons to hinder flock's alignment with the walls of funnels, or increase the size of the chevrons to promote the alignment.

\begin{figure}[hbt!]
\centering
\includegraphics[width=0.5\textwidth]{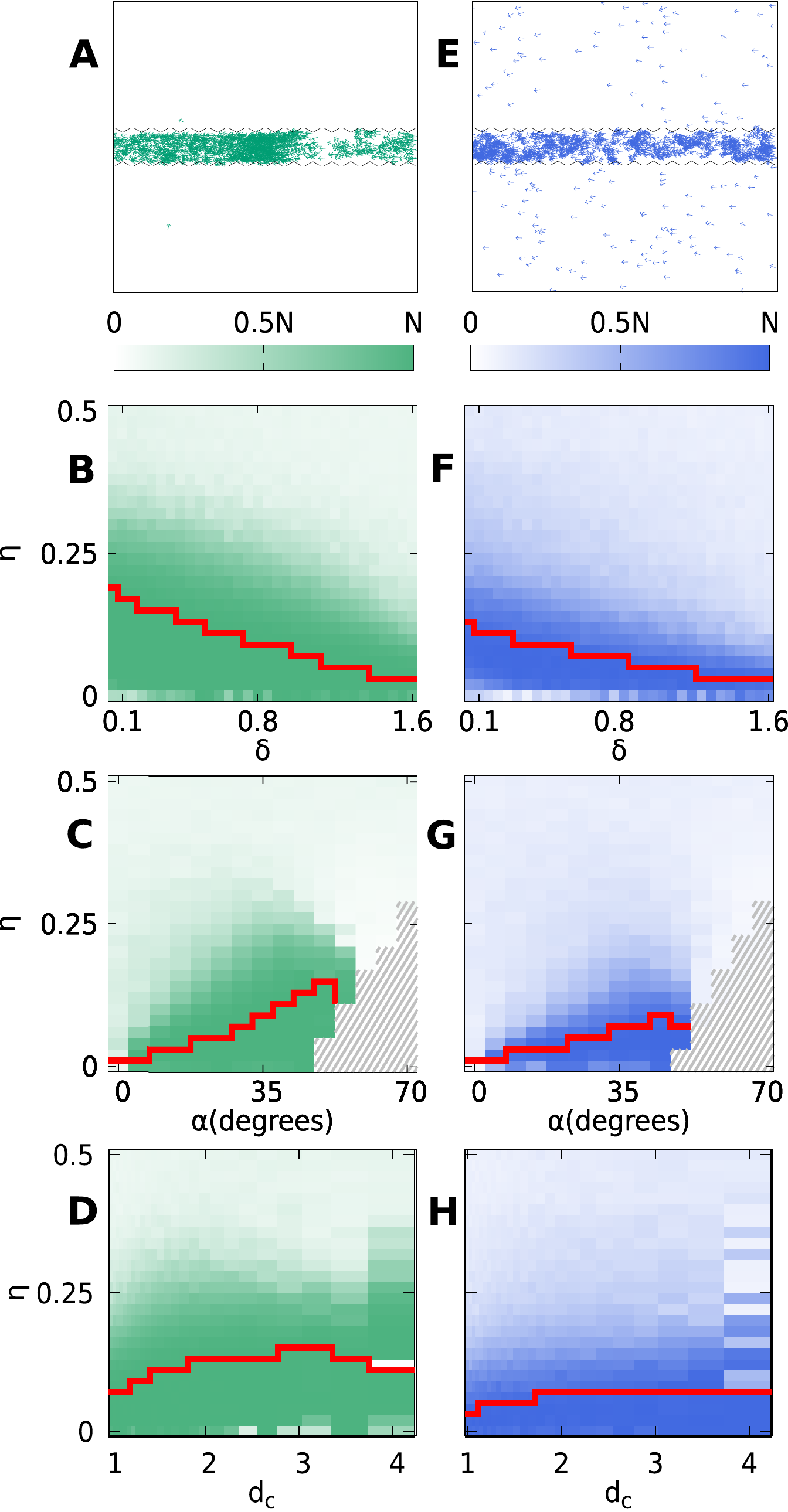}
\caption{{\bf Funnels geometry.} 
A (metric) E (topological):  Characteristic snapshot of the trapped steady state is shown for $\eta$ = 0.1, $d_c$= 2,
$\alpha$ = 30 degrees,  $\delta$ = 0.5 and A$_\text{trapping}$/A$_\text{total}$ = 1/10.
The colormaps represent the averaged number of trapped Vicsek particles over ten 
independent simulations as a function of $\eta$ and: $\delta$ with $d_c$ = 2 and $\alpha$ = 30 degrees 
for metric (B) and topological (F) Vicsek particles, $\alpha$ with $\delta$ = 1 and $d_c$ = 2 for metric (C)
and topological (G) Vicsek particles, and $d_c$ with $\delta$ = 0.5 and $\alpha$ = 30 degrees for metric (D)
and topological (H) Vicsek particles.
The red solid lines represent the threshold above which  dynamical trapping occurs and the 
grey dashed region corresponds to particles trapped inside the concave part of chevrons.
}
\label{etas}
\end{figure}

Next, we study the effect of both chevrons and funnels' geometry on the collective self-trapping of Vicsek particles.
To characterise this trapping behaviour, we reduce the area of the region in between the convex sides of the funnelled
walls to one tenth of the total area, by decreasing the distance between the two funnelled walls, thus forming a narrower
channel where particles tend to get trapped, as seen in Fig.~\ref{etas}A. Then, we establish the dependence of trapping 
on the opening length of the funnels $\delta$, the chevron angle $\alpha$, and the distance between neighbouring chevrons
$d_c$ (Fig.~\ref{comparison}A). 

A flock of Vicsek particles swimming along the concave side of the funnels is eventually pushed towards the convex
side as particles pass through the funnels openings $\delta$. Therefore, it is expected that $\delta$ plays a major
role in the collective self-trapping behaviour of Vicsek particles. For small funnels' openings, trapping within the region
in between the convex sides of the funnelled walls is effective up to high values of Vicsek noise $\eta$ (Fig.~\ref{etas}B). 
As $\delta$ increases, the probability of particles traveling from the convex towards the concave side increases,  
reason why self-trapping is effective up to lower values of $\eta$. We also explore the effect of the chevrons angle 
$\alpha$, while keeping $\delta$ and $d_c$ constant (Fig.~\ref{etas}C). For small angles there is a small difference
between the probability of particles passing towards the convex side from the concave one and viceversa, 
resulting into collective self-trapping only at small noises where particles form compact flocks aligned with the funnelled
walls. Similarly, chevrons with large angles only produce trapping at small noises since these funnels are not very effective
in keeping the flock aligned parallel to the wall. On the contrary, angles larger than 60 degrees promote the aligning of the
flocks perpendicular to the funnelled walls. Thus, the collective self-trapping of Vicsek particles occurs at an optimal trapping
of about 40 degrees. Finally, we study the effect of the distance between chevrons $d_c$, while keeping $\delta$ and $\alpha$
constant. As shown in Fig.~\ref{etas}D, trapping is not affected by $d_c$, and in all cases is not effective for noises greater
than 0.25.

We also consider the effect of the alignment rules between Vicsek particles and swimming velocity on this collective
trapping behaviour. We observe that elastic bouncing rules slightly enhance trapping with respect to the aligning rule,
but the overall behaviour remains unchanged (see Fig.~S3 in SM). However, the repulsive bouncing rule makes trapping 
a lot more erratic, and we no longer find a clear dependence with the opening length of the chevrons, $\delta$. Reducing 
the swimming speed, $v$, diminishes the importance of the bouncing rule in the trapping behaviour (see Fig.~S4 in SM).
At a swimming speed of $v=0.004$~\cite{drocco2012bidirectional}, the trapping of Vicsek particles is slightly more effective,
but the dependence with the opening length of the chevrons disappears. 

\begin{figure}[hbt!]
\centering
\includegraphics[width=0.5\textwidth]{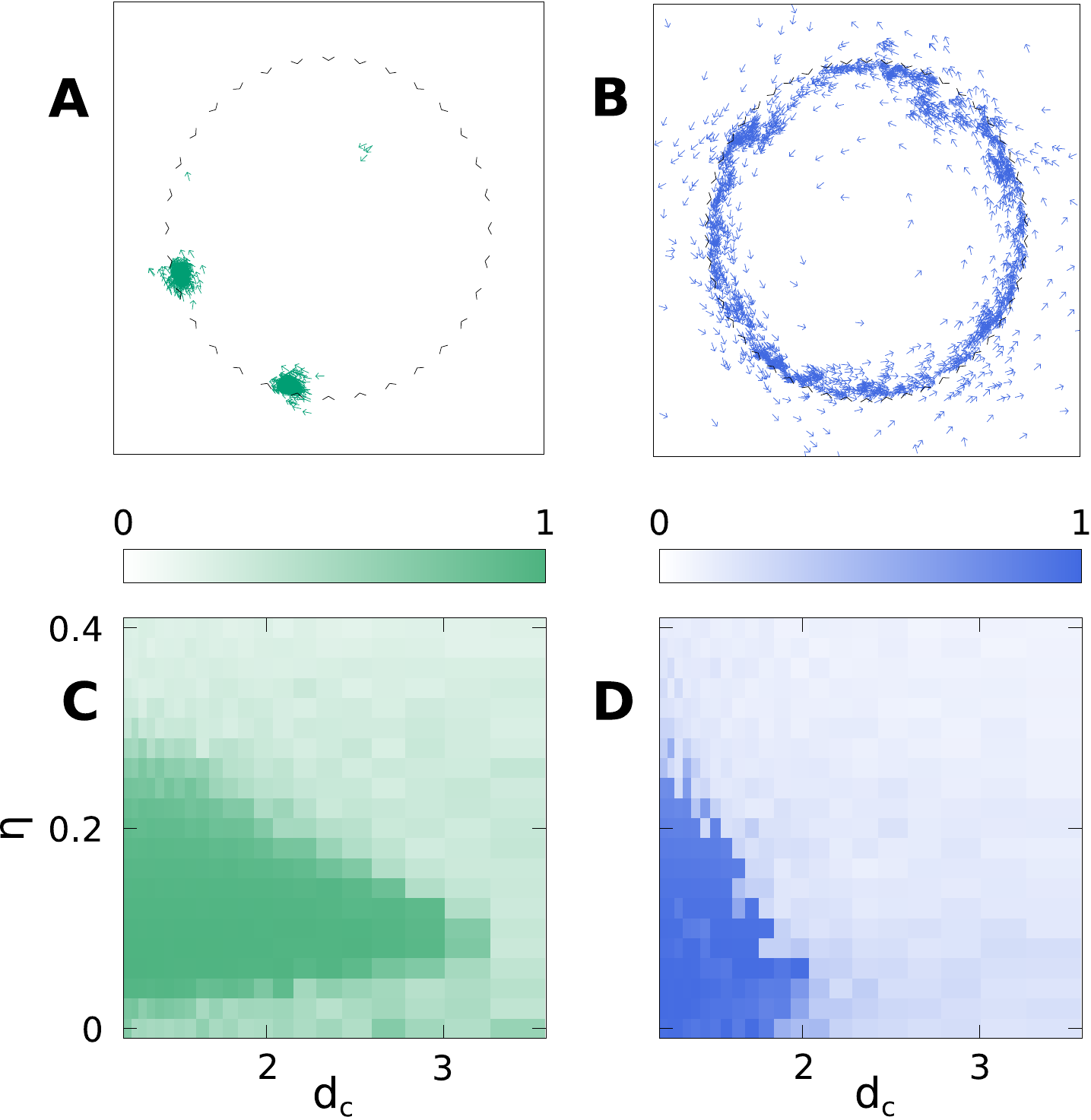}
\caption{{\bf Circular traps.} A) metric Vicsek particles get trapped in a circular trap made of funnels ($d_c = 2.356$).
B) Steady-state reached by topological Vicsek particles in the presence of a circular trap ($d_c = 1.508$). In both cases
the conditions where the same: $v=0.2$, $\delta=0.8$, $\alpha=30$ deg, radius of trap $R_t=12$ and 
C and D) Colormaps representing $\lambda$ as a function of $\eta$ and $d_c$. 
}
\label{circulos}
\end{figure}

Topological Vicsek particles exhibit the same type of collective self-trapping behaviour described for metric Vicsek
particles. However, due to their lower tendency to form flocks~\cite{ginelli2010relevance}, this collective trapping 
mechanism is less effective, resulting into a smaller number of particles trapped within the channel. Therefore, 
independently on the chosen parameter of the funnelled wall geometry, low noise levels are needed to enhance
the number of trapped particles as shown in Figs.~\ref{etas}F, G and H. The differences of the collective
self-trapping behaviour between metric and topological Vicsek particles is more evident when considering more
localised traps such as circular traps instead of infinite channels. These traps are formed by circular arrays of funnels
of radius $R_t$, whose convex side point towards the center of the trap, as shown in Fig.~\ref{circulos}.  Since we
apply periodic boundary conditions, this system is a square lattice of circular traps. Independently on the Vicsek model, 
either metric (Fig.~\ref{circulos}A) or topological (Fig.~\ref{circulos}B), Vicsek particles at low noise are preferentially
located inside the trap, while performing circular motion along the boundary. Metric and topological Vicsek differ in the
organisation of particles inside the trap, while metric Vicsek particles form flocks, topological ones are placed uniformly
along the circular perimeter. Since most of the particles are located at the circle's boundary, either inside or outside of
the funnelled wall, to quantify the trapping in this geometry instead of counting the number of particles inside the circle,
we calculate the number of particles performing a circular motion around the center of the trap by means of an order
parameter $\lambda\in [0,1]$ averaged over iterations in the steady state,
\begin{equation}
\label{lam}
\lambda =\frac{1}{N}\left|\sum_{i=1}^{N} \frac{(
(r_{xi}-r_{x0})v_{yi}-(
r_{yi}-r_{y0})v_{xi}}{\left| \vec{r}-\vec{r_0}\right|\left| \vec{v}\right|} \right|
\end{equation}
\noindent where $\vec{r}_0=(r_{x0},r_{y0})$ is the center of the circular trap, $\vec{r}_i=(r_{xi},r_{yi})$ and 
$\vec{v}_i=(v_{xi},v_{yi})$ are the ith-particle's position and velocity, respectively. When particles are trapped, thus 
performing a circular trajectory around its center, $\lambda$ is close to 1, while $\lambda$ will be lower if particles
are not trapped.  

In Figs.~\ref{circulos}C and D, the values of $\lambda$ as a function of the noise level and distance between
chevrons, $d_c$, are shown for both metric and topological Vicsek particles, respectively. In this case, the distance
between chevrons, $d_c$, is measured along the circumference arc. As it can be seen, the closer the chevrons,
the collective trapping within the circular trap becomes more effective. Although this trend is the same for both types
of Vicsek particles, metric Vicsek particles can be partially trapped, while topological Vicsek particles are either trapped
as seen in the snapshot of Fig.~\ref{circulos}B or particles are not trapped at all. Thus, metric Vicsek particles exhibit a
wider range of trapping inside the circles, but at the boundaries of this colormap we observe flocks that temporarily leave
a trap and eventually are confined by a neighbouring trap.

\begin{figure}[hbt!]
\centering
\includegraphics[width=0.95\linewidth]{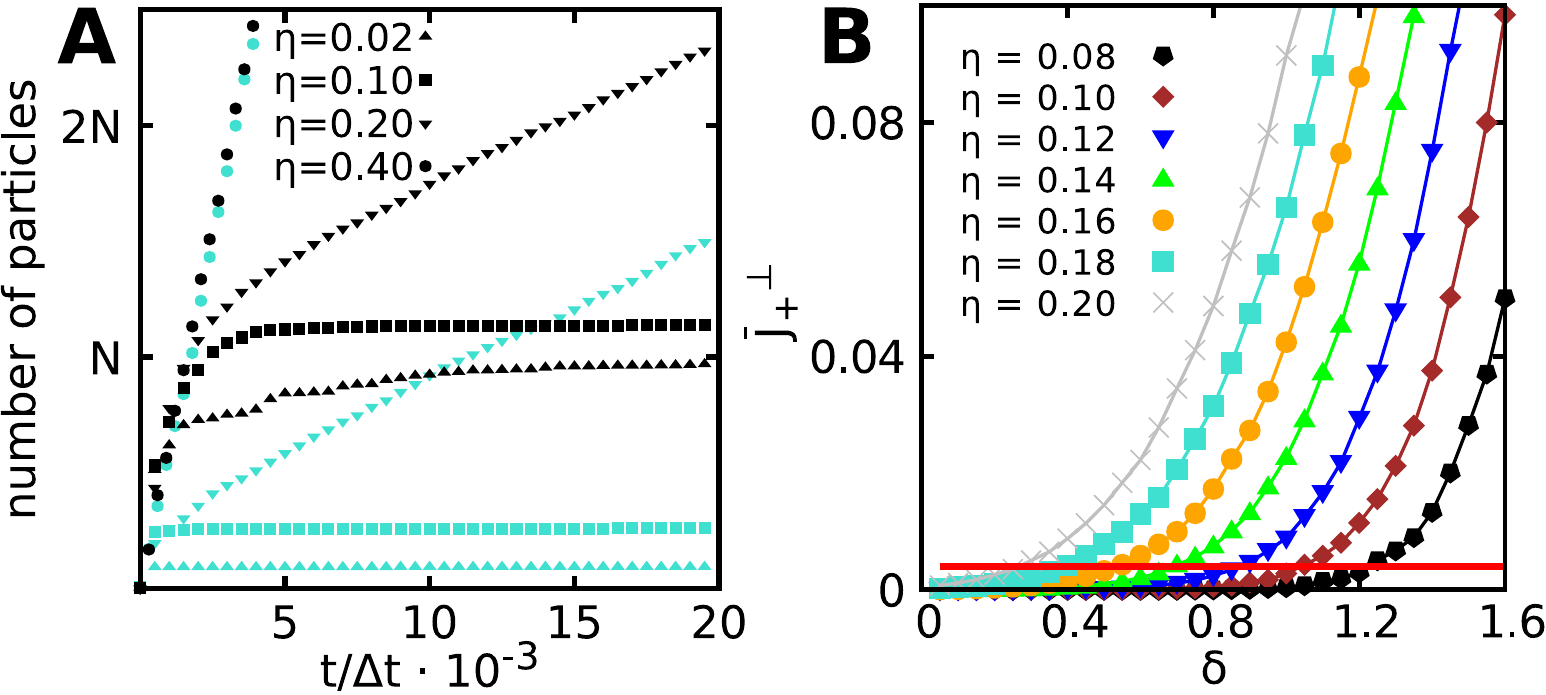}
\caption{{\bf Trapping dynamics.} A) Time evolution of the cumulative sum of particles transiting into the channel, 
represented by black symbols, and out of the channel, represented by light symbols at different Vicsek noises $\eta$: 
0.02 (up triangles), 0.1 (squares), 0.2 (down triangles) and 0.4 (circles). The chevron geometry is kept constant 
at $\delta$ = 0.3, $d_c$ = 2.0 and $\alpha$ = 30 degrees. 
B) Normalized flux of particles through the top walls of the channel, $\bar{J}^{\perp}_{+}$, as a function of the
opening between chevrons, $\delta$, for several values of Vicsek noises, 0.08 (black pentagons), 0.10 (brown diamonds), 
0.12 (blue down triangles), 0.14 (green up triangles), 0.16 (orange circles), 0.18 (light blues squares) and 0.20 
(grey crosses). The horizontal red line corresponds to the threshold value above which we consider a dynamical
trapping of Vicsek particles.
}
\label{dyna}
\end{figure} 


We now study the dynamics of the collective trapping of Vicsek particles in channels made of funnels(Fig.\ref{dyna}).
For a given chevron geometry we compute the time evolution of  particles transiting through the funnelled walls in 
(dark symbols) and out of the channel (light symbols). As shown in Fig.\ref{dyna}A, at low Vicsek noises (up triangles),
flocks form and more particles enter into the channel than particles leave it, although the system does not reach steady
state during the simulated time. As the noise increases, $\eta$ = 0.1 (squares), the rate of particles entering the channel
is higher, reaching a trapping steady state at shorter times. For $\eta$ around 0.2 (down triangles) the system reaches
a different steady state in which particles are constantly passing through the funnelled walls in both directions, while keeping
most of the particles inside the channel. We refer to this behaviour as dynamical trapping. At higher noises, above 
$\eta$ = 0.4 (circles), Vicsek particles do not experience funnels' anisotropies,  resulting in a homogeneous distribution
of particles.

The number of particles crossing  from the convex side to the concave side of the channel through the top funnelled wall,
$N^{\perp}_+$, and bottom wall, $N^{\perp}_-$, results into a perpendicular flux of particles escaping
the channel through the top wall at each time step, J$^{\perp}_+ = \frac{N^{\perp}_+}{L \Delta t}$ or the bottom wall, 
J$^{\perp}_- = \frac{N^{\perp}_-}{L \Delta t}$, respectively. The normalised flows through the top wall, 
$\bar{J}^{\perp}_{+}$  = J$^{\perp}_{+} \frac{1}{\rho v}$, as a function of the opening between chevrons for different 
Vicsek noises are shown in Fig.~\ref{dyna}B. For small values of $\delta$, Vicsek particles are trapped within the channel
and thus, there is not flux of particles through the walls. As $\delta$ increases, particles are able to escape from the channel
and the flux of particles through the walls increases. The value of $\delta$ at which particles start to escape from the channel
decreases with the Vicsek noise. To define the boundary between trapping and dynamical trapping regimes, we arbitrarily
choose a minimum flux of particles through the top wall, $\bar{J}^{\perp}_{+}$ = 0.004 (horizontal red line in Fig.~\ref{dyna}B).
The boundary between these two trapping regimes is shown by the solid red lines in Fig.~\ref{etas} as a function of the funnel's
geometry ($\delta$, $d_c$ and $\alpha$) and Vicsek noise ($\eta$). We observe the same dynamical regimes in the topological
Vicsek model, but at smaller Vicsek noises (see Fig.~S6 in the SM).

\begin{figure}[hbt!]
\centering
\includegraphics[width=0.95\linewidth]{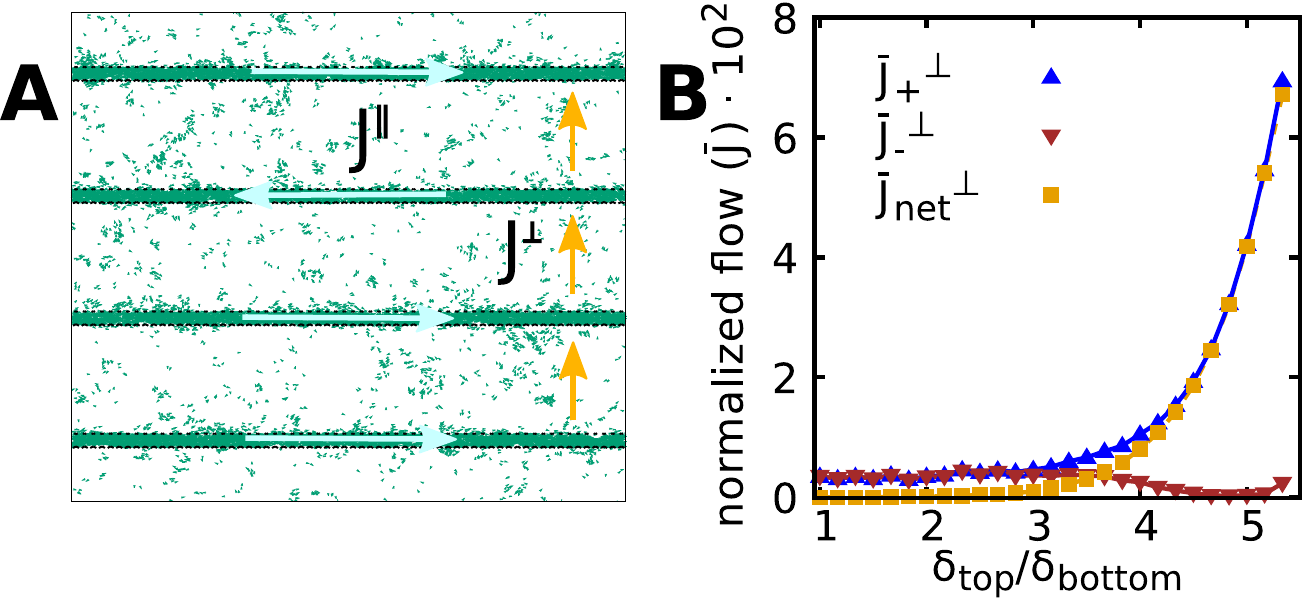}
\caption{{\bf Cross flow of metric Vicsek particles.} A) Snapshot of the simulation box containing four parallel channels
formed by pairs of oppositely oriented funnelled walls, with the bottom wall with opening of $\delta_{bottom}$ = 0.3
and the top wall has an opening of $\delta_{top}$ = 1.2  ($d_c$ = 2, $\alpha$ = 30 degrees,  $\eta$ = 0.12, $L$  = 128, 
$N$ = 32000 and $v$ = 0.2). The light blue arrow shows the swimming direction of each flock along the channels and the
orange the perpendicular direction. B) Normalised flows of Vicsek particles: net perpendicular flow $\bar{J}^{\perp}_{net}$ 
(orange squares), top perpendicular flow $\bar{J}^{\perp}_+$  (blue top triangles) and down perpendicular flow 
$\bar{J}^{\perp}_-$ (red down triangles).
 }
\label{flows}
\end{figure}

Combining the dynamical trapping regime with different geometries of the top and bottom funnelled walls of a channel,
we can obtain two perpendicular flows of Vicsek particles. Instead of a single channel, we simulate four parallel channels
to minimise possible finite size effects (Fig~\ref{flows}A). The bottom wall of the channels has an opening of
$\delta_{bottom}$ = 0.3, while the top wall has wider openings ranging from $\delta_{top}$ = 0.3 to $\delta_{top}$ = 1.6.
Thus, the bottom wall works in the trapping regime, while the top wall works in the dynamical trapping regime except in 
the symmetric case, $\delta_{top}$ = $\delta_{bottom}$, where the system is in the trapping regime and
Vicsek particles are trapped within the channels: $\bar{J}^{\perp}_{+} = \bar{J}^{\perp}_{-} \approx 0$ (Fig~\ref{flows}B). 
As soon as we consider the top wall working in the dynamical trapping regime, $\delta_{top}/\delta_{bottom}$ = 3, the
Vicsek particles have a higher probability to pass through the funnels with wider openings, a fraction of which travels out
of the channel towards the neighbouring channel. This results into a net flux of Vicsek particles along the perpendicular
direction of the channels towards the top region, $\bar{J}^{\perp}_{net}=\bar{J}^{\perp}_{+}-\bar{J}^{\perp}_{-}$. 
This flux perpendicular to the channel increases as the opening of the funnels of the top wall increases, $\delta_{top}$, 
as shown in Fig.~\ref{flows}. Even though each channel is also characterised by a net flux of particles along the channel
(J$^{\vert\vert}$), there is not net flux for the entire system due to the absence of a preferential swimming direction parallel
to the channels. However, a net flux of particles parallel to the channels could be obtained by breaking the symmetry along
the channels either by tilting the chevrons or adding a chevron inside of the channel. 

\begin{figure}[hbt!]
\centering
\includegraphics[width=0.5\textwidth]{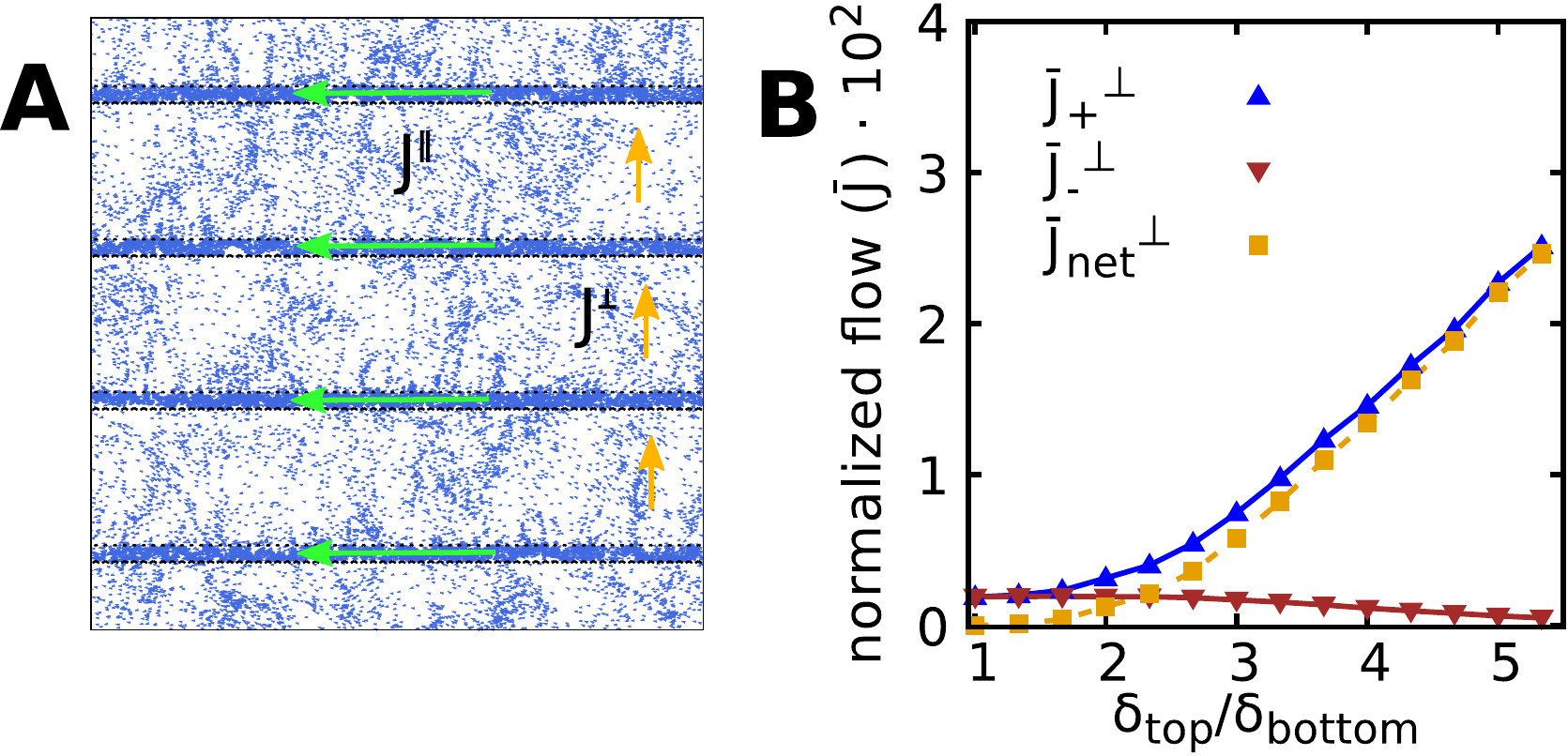}
\includegraphics[width=0.5\textwidth]{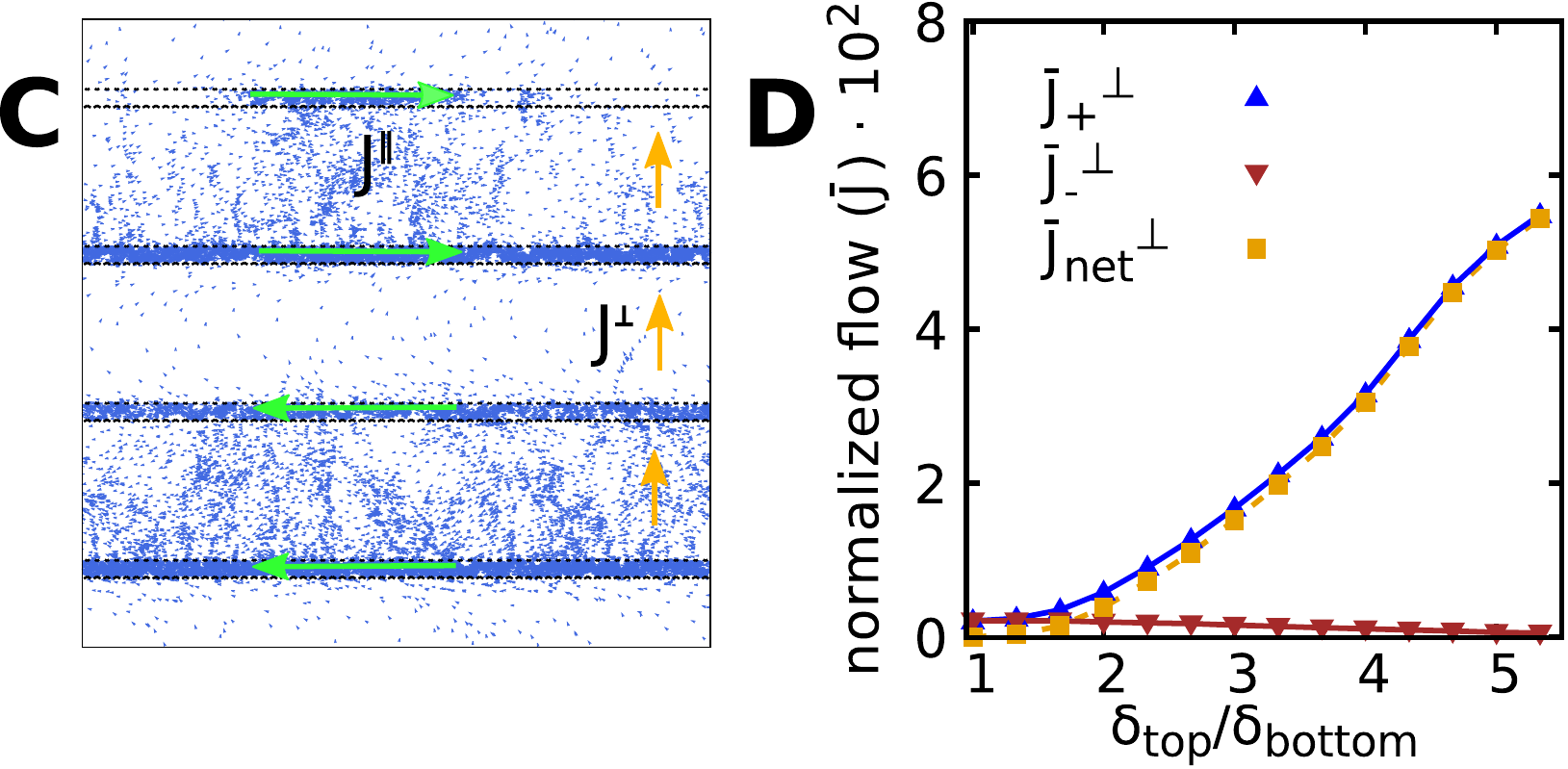}
\caption{ {\bf Cross flow of topological Vicsek particles.} Top row, a snapshot of a situation where all flows have
the same direction and the results for $J_{\perp}$ and $J_{\vert\vert}$ in that case. Bottom row, the same for the opposite
flows situation. All parameters have been taken to be the same than in figure 4 of the main text, but using topological 
Vicsek model instead of metric.
}
\label{tft}
\end{figure}

The same behaviour is observed for topological Vicsek particles within channels formed by a funnelled wall 
working on the trapping regime combined with a wall working on the dynamical trapping regime. However, while for
metric Vicsek particles the direction of the flocks swimming along the channel does not have any effect on the magnitude
of the perpendicular flows of Vicsek particles, in the case of topological Vicsek particles the relative direction of the
swimming direction of the flocks between two adjacent channels strongly affects the magnitude of the perpendicular
flows of Vicsek particles, $\bar{J}^{\perp}$. In Fig.~\ref{tft}A we show a snapshot of the simulation box when all four
channels have their Vicsek flows aligned. The corresponding cross flows measured as a function of the wider $\delta$
value are shown in Fig.~\ref{tft}B, which are similar to those obtained for metric Vicsek particles. In Fig.\ref{tft}C and D,
we show a system where there are Vicsek flows along the channels in different directions and the dependence of the
cross flows with $\delta$ using only cases with opposite flows, respectively. As it can be seen, when the Vicsek flows
between two neighbouring channels have opposite directions, the cross flows are considerably enhanced. Since topological
Vicsek particles interact with their first Voronoi neighbours, particles that scape from one bottom channel gradually change
their velocities to align it with that of the flock in the top channel. This effectively push particles towards the top channel,
reducing the density of particles in the region in between two channels with flocks swimming in opposite directions,
as shown in Fig.~\ref{tft}C. Given that the definition of each particle neighbours does not depend on the distance,
this interaction is independent of the distance between channels and thus, the cross flow in this asymmetric case 
only depends on the phase between the channels, which is constant at 180 degrees.

\section{Conclusions}

To summarise, we have studied the trapping behaviour of active particles by means of anisotropic obstacles.
V-shaped obstacles induce the rectification of active particles, resulting into the concentration of run-and-tumbling
particles within the region at the concave side of channels formed by oppositely oriented walls of funnels.
On the contrary, particles interacting via aligning interactions, as described by the Vicsek model, concentrate at the 
convex side of the funnelled walls. Differently from run-and-tumbling particles, the trapping behaviour of Vicsek particles
is driven by collective effects. We have shown that this collective trapping is independent of the aligning rules with the
obstacles, either repulsive, elastic or aligning. In addition, this effect is not dependent on the definition of particles's neighbours,
either metric or topological Vicsek particles. The key element to observe this collective self-trapping behaviour is the alignment
of the flocks of particles with the funnelled walls; systems in which flocks are not allowed to align with the walls of funnels
tend to concentrate on the concave side of the walls, similar to run-and-tumbling particles. We carefully considered how the
geometry of the wall of funnels condition the collective trapping behaviour exhibited by Vicsek particles. We observe that
trapping is more effective for narrow openings of the funnels and short distances between chevrons when
considering an optimal angle of the chevrons of about 40 degrees. Moreover, we observed three well-defined 
dynamical regimes for this collective trapping effect as a function of the geometry of the funnels and Vicsek noise:
i) particles pass through a wall of funnels from the convex side towards the concave side of the wall,
resulting into their trapping within this region of the channel between the funnel's walls, ii) particles pass from both
sides of a wall of funnels at equal rates and do not get trapped and iii) particles exhibit different rates passing through
a wall funnels from one side or the other, reaching a steady state in which particles are effectively trapped within the
channel, but are constantly transiting through the walls of funnels. Exploiting the latter dynamical trapping regime
we engineered a system that exhibits two perpendicular flows of Vicsek particles, whose magnitude and direction
can be tuned by modifying the geometry of the walls of funnels.    

\section{Acnowledmennts}

The authors acknowledge funding from Grant FIS2016-78847-P of the MICINN and the UCM/Santander PR26/16-
10B-2. F.A. acknowledges support from the Juan de la Cierva program (FJCI-2017-33580). 
The project that gave rise to these results received the support of a fellowship from ''la Caixa" Foundation (ID 100010434). The fellowship code is LCF/BQ/LI18/11630021. R.M. thanks MICINN by the award of a FPI pre-doctoral grant (BES-2017-081108) and J.L.A acknowledges the support of the IFIMAC (MDM-2014-0377) and NVIDIA corporation
with the donation of the Titan Xp GPU used for this research.

\bibliography{referencias.bib}

\end{document}